\documentclass[useAMS,usenatbib]{mn2e}
\usepackage{epsfig}
\usepackage{natbib}

\DeclareMathVersion{bold}

\title{Statistical analysis of undetected point sources in cosmic microwave background maps}

\author[Arg\"ueso et al.]{F. Arg\"ueso$^{1}$ \footnotemark,
J. L. Sanz$^{2}$, R.B. Barreiro$^{2,3}$, D. Herranz$^{2}$, J.
Gonz\'alez-Nuevo$^{4}$
\\  $^{1}$ Departamento de Matem\'aticas, Universidad de Oviedo, Avda. Calvo Sotelo s/n, 33007 Oviedo, Spain
\\  $^{2}$ Instituto de F\'\i{sica} de Cantabria (CSIC-UC), Avda. los Castros  s/n, 39005 Santander, Spain
\\  $^{3}$ Departamento de F\'\i sica Moderna, Universidad de Cantabria, Avda. los Castros, s/n, 39005 Santander, Spain
\\  $^{4}$ SISSA-I.S.A.S, via Beirut 4, I-34014 Trieste, Italy}

\begin{document}

\maketitle

\begin{abstract}
Cosmic microwave background (CMB) temperature anisotropies follow a
Gaussian statistical distribution in the standard inflationary
model, but there are non-Gaussian contributions due to astrophysical
foregrounds. The detection of the non-Gaussian signal due to
extragalactic point sources and its distinction from the possible
intrinsic non-Gaussianity is an issue of great importance in CMB
analysis. The Mexican Hat Wavelet Family (MHWF), which has been
proved very useful for the detection of extragalactic point sources,
is applied here to the study of non-Gaussianity due to point sources
in CMB maps. We carry out simulations of CMB maps with the
characteristics of the forthcoming Planck mission at 70 and 100 GHz
and filter them with the MHWF. By comparing the skewness and
kurtosis of simulated maps with and without point sources, we are
able to detect clearly the non-Gaussian signal due to point sources
for flux limits as low as 0.4 Jy (70 GHz) and 0.3 Jy (100 GHz). The
second and third members of the MHWF perform better in this respect
than the Mexican Hat Wavelet (MHW1) and much better than the
Daubechies 4 wavelet. We have also estimated the third order, $K_3$,
and fourth order, $K_4$, cumulants produced by point sources at
these Planck channels by means of a fit with the MHWF. The average
relative errors with respect to the real values are below $12\%$ for
fluxes down to $0.6$ Jy (70 GHz) and $0.4$ Jy (100 GHz). The values
of these cumulants allow us to distinguish between different source
counts models. From the estimated cumulants and assuming a power law
for the source number counts we are able to obtain the coefficients
$\alpha$ and $A$ of the differential number counts, $\alpha=2.19\pm
0.46$ ($\alpha=2.26\pm 0.19$), $A=24.3\pm 4.3$ ($A=21.0\pm 4.2$) for
70 (100) GHz, assuming a flux limit of 1 Jy. These results are
consistent with the values obtained from simulations containing only
point sources, which are: $\alpha=2.32\pm 0.06 $ ($\alpha=2.35\pm
0.16$), $A=22.1\pm 1.5$ ($A=20.0\pm 3.7$) for 70 (100) GHz.

\end{abstract}

\begin{keywords}
methods:statistical-cosmic microwave background
\end{keywords}

\section{Introduction}
\footnotetext{E-mail: argueso@uniovi.es}

According to the standard cosmological model, CMB temperature
anisotropies are caused by the inhomogeneities in the distribution
of matter and radiation present at the decoupling time. These
inhomogeneities are the product of quantum fluctuations amplified in
the inflationary era and follow a Gaussian statistical distribution.
Therefore, the observed CMB anisotropies are a realization of a
homogeneous and isotropic Gaussian random field on the sphere.
Nevertheless,  some degree of non-Gaussianity can be introduced, for
instance, by non standard inflation \citep{bar04} or by topological
defects \citep{tur90,dur99}. The analysis of the first and
three-year Wilkinson Microwave Anisotropy Probe (WMAP) data by the
WMAP team shows that CMB temperature fluctuations are consistent
with a Gaussian distribution \citep{kom03,spe06} in agreement with
the standard inflation paradigm. However, some studies have detected
non-Gaussian features in the WMAP data, for instance in \cite{vie04}
and \cite{cru05,cru06} a cold non-Gaussian spot of unknown origin
was detected and studied by using the Spherical Mexican Hat Wavelet.
Non-Gaussian signatures have also been found by using different
methods: phase correlations \citep{chi03,col04}, local curvature
methods \citep{han04,cab05}, multivariate analysis \citep{din05},
Minkowski functionals \citep{par04,eri04} and higher criticism
\citep{cay05}.

On the other hand, astrophysical foregrounds contaminate the
cosmological signal and also give rise to a non-Gaussian
contribution. In particular, the emission due to point sources
(radio and infrared galaxies are seen as point-like objects due to
their very small projected angular size in comparison with the
typical experimental resolutions) is clearly non-Gaussian
\citep{kom01,kom03,arg03,pie03,her04,bar06}. In the last few years,
new high frequency ($\nu\geq 10$ GHz) surveys of radio sources have
been carried out \citep{wal03,ric04}. All the recently published
data on counts and on the emission properties of extragalactic radio
sources have guided the development of new cosmological evolution
models able to improve on previous ones \citep{dun90,tof98} in
predicting number counts and source statistics at high radio
frequencies \citep{dez05}. We will use this new model to carry out
point source simulations.

It is very important to assess the degree of non-Gaussianity
generated by point sources in order to distinguish it from the
non-Gaussian signal coming from other origins. The ESA Planck
satellite \citep{man98,pug98}, which will be launched in 2008, will
analyze the CMB anisotropies with unprecedented accuracy and
resolution and will make possible a very precise non-Gaussianity
study, for which the use of wavelets can be very suitable.

 The Mexican Hat Wavelet (MHW1) has been very useful in detecting point
sources \citep{cay00,vie01a,vie01b} in CMB maps, by using the signal
amplification going from real space to wavelet space. The Mexican
Hat Wavelet Family (MHWF) was introduced by \cite{gon06} as an
extension of the standard MHW1, which produced higher amplification
(see below) and then a better performance at point source detection.
These new wavelets were also used for point source detection in
Planck maps in \cite{lop06}. The MHWF is a family of isotropic 2D
wavelets obtained by applying the Laplacian operator iteratively to
the 2D Gaussian.

The expression of these new wavelets in real space is
\begin{equation}
\label{cc} \psi_{n}(x)=\frac{(-1)^n}{
2^{n}n!}\triangle^{n}{\varphi(x)}
\end{equation}
\noindent where $x$ is the radial coordinate and $\varphi$ is the 2D
Gaussian
\begin{equation}
\label{cc}
 \varphi(x)=\displaystyle\frac{e^{-x^2/2}}{2\pi}
  \end{equation}

\noindent   Note that $\psi_{1}(x)$ is the standard MHW1, we call
the other corresponding members of the family MHWn. Any member of
the family can be written very easily in Fourier space as
\begin{equation}
\hat\psi_{n}(k)=\frac{k^{2n}e^{-\frac{k^{2}}{2}}}{2^{n}n!}
\end{equation}

\noindent \cite{gon06} considered a point source convolved with a
Gaussian instrumental beam of dispersion $\gamma$
\begin{equation}
\gamma=\frac {FWHM}{\sqrt{8\log(2)}}
\end{equation}
\noindent where FWHM denotes the Full Width Half Maximum of the
instrumental beam.

The point source was embedded in a CMB map, including noise and all
the relevant foregrounds. Then, they filtered the map with the first
members of the MHWF at different scales R.
\begin{equation}
 \Psi_{n}(\vec{x}; \vec{b}, R) \equiv \frac{1}{R^{2}}\psi_{n}
\Big(\frac{|\vec{x}-\vec{b}|}{R}\Big),
\end{equation}

\noindent The amplification, $\lambda$, is defined as the ratio
between the intensity of the point source $I_w$ and the rms
deviation $\sigma_w$ in the wavelet filtered image divided by this
ratio in the original image filtered with a Gaussian-shaped
instrumental beam.

\begin{equation}
\lambda=\frac{I_{w}/\sigma_w}{I_{g}/\sigma_{g}}
\end{equation}

\noindent $I_w$ can be calculated for the different members of the
MHWF as

\begin{equation}
\label{cc}
 I_{w}=\displaystyle\frac{I_g \beta^{2n}}{(1+\beta^{2})^{n+1}}
\end{equation}

 \noindent with $\beta=R/\gamma$.

 It was proved that when filtering CMB maps with
the first three members of the MHWF at an optimal scale, \textit{R},
the MHW2 and MHW3 produced a higher amplification than the standard
MHW1 and were hence more effective at source detection. Other higher
n members of the MHWF did not yield higher amplifications.

After removing the detected point sources, there could still be a
remaining non-Gaussianity due to undetected point sources. It is of
great interest to characterize this non-Gaussianity in order to
distinguish it from the intrinsic one.

The detection of the non-Gaussian signal due to point sources poses
a similar problem to that of point source detection, these new
wavelets will enhance the signal (non-Gaussian) due to point
sources, thus making it possible to detect this kind of
non-Gaussianity more effectively. Since the skewness and kurtosis of
a Gaussian distribution are zero and three respectively, any
significant deviation from these values will mean the presence of a
non-Gaussian signal, so that the skewness and the kurtosis of the
wavelet filtered maps can be used as estimators of the
non-Gaussianity due to point sources.

Filtering CMB maps with the MHWF at optimal scales amplifies point
sources and can also be used to determine the third and fourth order
cumulants produced by the point sources. From these cumulants, we
can also determine the parameters $A$ and $\alpha$, if we assume
that the differential number counts can be expressed as a power law
$dn/ds=AS^{-\alpha}$. In \cite{pie03} a chi-squared fit to a power
law was explored considering the moments of the CMB maps. We have
considered in this paper a chi-squared fit of the third order and
fourth order cumulants of the wavelet filtered maps combining
several scales including the optimal ones.

The plan of the paper is as follows: in \S 2, we calculate the
cumulants of a wavelet filtered map from those of the original map,
this allows us to relate the skewness and kurtosis of the filtered
maps to the amplification. In \S 3, we carry out realistic CMB
simulations with and without point sources, considering the
characteristics of the Planck experiment at 70 and 100 GHz and
perform the non-Gaussianity detection by filtering with the MHWF. In
\S 4, we obtain the third and fourth order cumulants produced by
point sources by means of a fit of the cumulants of wavelet filtered
CMB maps and we also determine the parameters $\alpha$ and $A$.
Finally in \S 5, we draw the main conclusions of the paper.

\section{Skewness and kurtosis of a wavelet filtered map}

Skewness and kurtosis provide us with suitable tests for determining
whether certain data come from a Gaussian distribution or not. The
skewness, $\textit{s}$,  is defined as
\begin{equation}
\textit{s}=\displaystyle\frac{\mu_3}{\sigma^3}
\end{equation}

\noindent where $\mu_3$ is the third central moment of the
distribution and $\sigma$ its rms deviation, the skewness is zero
for a Gaussian distribution. The kurtosis, \textit{k},  is the
fourth central moment divided by the squared variance

\begin{equation}
\textit{k}=\displaystyle\frac{\mu_4}{\sigma^4}
\end{equation}

\noindent and its value is 3 in the Gaussian case. We are interested
now in calculating the skewness and kurtosis due to point sources in
a wavelet filtered map from these quantities in the original
observed map. Then we consider a pixelized flat 2D image including
CMB, Galactic foregrounds, instrumental noise and extragalactic
point sources; these four components are independent and are added
up to form the image; whereas the CMB is Gaussian and the noise is
considered as non-uniform but Gaussian-distributed at each pixel,
the other two are non-Gaussian.

Skewness and kurtosis can be obtained from the cumulants of the
probability distribution function (pdf) $ p(x)$ . The characteristic
function $ \phi (u)$ is defined as the Fourier transform of the pdf
\begin{equation}
\phi(u)=\int p(x)\,e^{iux}\, dx
\end{equation}

\noindent and the cumulant function is $ K(u)= log(\phi(u))$. The
cumulants of order m,  $K_m$, can be calculated as derivatives of
the cumulant function
\begin{equation}
K_m=\frac{1}{i^m} \frac {d^{m}K(u)}{du^m}\big|_{u=0}
\end{equation}

 The second order cumulant is the variance, the third order
cumulant is the third central moment $\mu_3$ and the fourth order
cumulant is $\mu_4-3\sigma^4$. The advantage of using the cumulants
is that when the different components of the data are independent
the cumulant of the sum is the sum of the cumulants. So, the
cumulant of any order of the sum of CMB, noise, point sources and
other foregrounds is the sum of the cumulants of the components.

In a typical CMB experiment, the image is filtered with a
Gaussian-shaped instrumental beam, this means that the map is
convolved with a Gaussian function. This convolution changes the
skewness and kurtosis of the original map. After the convolution,
the temperature of a pixel is a linear combination of the
temperatures of the neighboring pixels in the unfiltered map. For
point sources, which are not correlated (this is true for the
frequencies, 70 and 100 GHz, considered in this work, as can be seen
in \cite{gon05}), the cumulants of this linear combination can then
be written as a linear combination of the cumulants of the original
map. If we are filtering with a function $\varphi(x)$, the cumulant
of order m, $K^f_m$, of the filtered image can be written as

\begin{equation}
\label{c6}
 K^f_m=K_m A^{m-1}_p \int \varphi^m(\vec{x})\, d\vec{x}
\end{equation}

\noindent where $K_m$ is the original cumulant of order m in the
pixelized image and $A_p$ the pixel area. From this formula, we can
obtain the cumulants of the Gaussian filtered image $ K_{m}^{g}$
from those of the original one.

\begin{equation}
K_{m}^{g}=K_{m}\frac{1}{m(2\pi)^{m-1}}\left(\frac{l}{\gamma}\right)^{2m-2}
\end{equation}

\noindent where $l$ is the pixel size and $\gamma$ the beam
dispersion, see eq. (4). We are going to filter now the image with
the first three members of the MHWF, the whole process amounts to
filtering the original image with a convolution of the Gaussian
instrumental beam and the corresponding wavelets. We could then
apply Eq.(12) with $\varphi(\vec{x})$ this convolution. Finally, we
can relate the cumulants $ K_{m}^{w} $ of the wavelet filtered image
to the same quantities $ K_m$ in the original pixelized image. We
write the formula for a general cumulant and a general member of
index n of the MHWF

\begin{equation}
K_{m}^{w}=K_{m} \frac{2\pi
R^{2-2m}C_{mn}}{l^{2-2m}(1+\beta^{-2})^{(n+1)m-1}}
\end{equation}

\noindent in this formula $\beta=R/\gamma$, R is the wavelet scale
(see eq. 5) and the coefficients $C_{mn}$ are calculated as

\begin{equation}
C_{mn}=\int_{0}^{\infty} (\psi_{n}(x))^{m}\, x\, dx
\end{equation}

\noindent This last expression can be easily computed for the first
members of the MHWF and for m=3,4.

When we consider the skewness of the total wavelet filtered image
(CMB+point sources+Galactic foregrounds+noise), it can be obtained
as the total third order cumulant divided by $\sigma_w^3$ with
$\sigma_w$ the total rms deviation. The contribution of point
sources to the total skewness of the wavelet filtered map is

\begin{equation}
\label{cc}
 \hat {s}_w=\frac{K^{w}_{3}} {\sigma^3_w}
\end{equation}

\noindent where $ K^w_3$ is the third order cumulant (due to point
sources) of the wavelet filtered map. The contribution of point
sources to the total skewness of the Gaussian filtered map is

\begin{equation}
\label{cc}
 \hat {s}_g=\frac{K^{g}_{3}} {\sigma^3_g}
\end{equation}

\noindent where $ K^g_3$ is the third order cumulant (due to point
sources) of the Gaussian filtered map and $\sigma_g$ the total rms
deviation of the Gaussian filtered maps. Finally, by using Eqs. (13)
and (14) and the definition of the amplification, $\lambda$, we can
relate $\hat{s}_w$ and $\hat{s}_g$. We obtain the following formula
for the MHW1

\begin{equation}
\label{cc}
 \hat {s}_w=\frac{4}{9} (1+\beta ^2)\lambda^3 \hat{s}_g
\end{equation}

\noindent for the MHW2, we obtain a similar formula

\begin{equation}
\label{cc}
 \hat {s}_w=\frac{22}{81} (1+\beta ^2)\lambda^3 \hat{s}_g
\end{equation}

\noindent and finally for the MHW3

\begin{equation}
\label{cc}
 \hat {s}_w=\frac{1240}{6561} (1+\beta ^2)\lambda^3 \hat{s}_g
\end{equation}

\noindent We want to point out that these contributions depend on
the cubed amplification and will be higher for the wavelets with the
highest amplifications, this suggests that MHW2 and MHW3 can be
better at detecting non-Gaussianity due to point sources than MHW1.
Besides, we will choose optimal scales when filtering with the MHWF
so that the contribution to the skewness is maximum.

We can also obtain similar relations between the contribution of
point sources $ \hat {k}_w $ to the kurtosis of the wavelet filtered
maps and the contribution of point sources $ \hat {k}_g $ to the
kurtosis of the beam filtered maps. So, we have for the MHW1

\begin{equation}
\label{cc}
 (\hat {k}_w-3)=\frac{15}{32} (1+\beta ^2)\lambda^4 (\hat{k}_g-3)
\end{equation}

\noindent for the MHW2

\begin{equation}
\label{cc}
 (\hat {k}_w-3)=\frac{2547}{8192} (1+\beta ^2)\lambda^4
 (\hat{k}_g-3)
\end{equation}

\noindent and for the MHW3

\begin{equation}
\label{cc}
 (\hat {k}_w-3)=\frac{61731}{262144} (1+\beta ^2)\lambda^4
 (\hat{k}_g-3)
\end{equation}

\noindent In this case, the contribution of point sources to the
total kurtosis in the wavelet filtered map is proportional to
$\lambda^4$.

\section{Non-Gaussianity of simulated CMB maps}

In this section, we are going to consider the study of the
non-Gaussianity in realistic simulated CMB maps. We have carried out
simulations of CMB flat 2D maps of $12.8\times12.8$ square degrees,
generated with the cosmological parameters of the standard model
\citep{spe03}. We have then added the relevant Galactic foregrounds
(free-free, synchrotron and dust emission), we have used the Planck
Reference Sky, prepared by the members of the Planck Working Group
2. Galactic foregrounds have been selected from 12 different regions
of the sky (at Galactic latitude $b>30^\circ$). Each one of these 12
Galactic simulations has been used according to its
representativeness as in \cite{lop06}.

  In order to give the most realistic numbers of detected
extragalactic point sources, we have adopted the model recently
presented by \cite{dez05}. This new model takes into account the new
data coming from high frequency radio surveys of extragalactic
sources and discusses all source populations contributing to the
number counts at these frequencies. It has thus proven capable of
giving a better fit than before to all the currently published
source number counts and related statistics at $\nu\geq 10$ GHz
coming from different surveys \citep{ric04,wal03}. This model also
fits well the number counts obtained from the three year WMAP data
\citep{hin06}. 

Since we are especially interested in applying our new method to the
maps which will be provided by the {\it Planck} mission in the near
future, and given that we can be very confident in the input source
model counts, we have considered the specific conditions of two {\it
Planck} channels, which operate at 70 and 100 GHz.\footnote{The
characteristics of the channels, relevant for our purposes, are: a)
pixel sizes, $3'$ and $3'$  at 70 and 100 GHz, respectively; b) Full
Width Half Maximum (FWHM) of the circular gaussian beams, $14'$ and
$9.5'$, respectively. In all the cases, we have used the estimated
instrument performance goals available at the web site:
$http://astro.estec.esa.nl/Planck$.}

We are going to consider that the noise is anisotropic. Anisotropic
noise has been generated using the simulation pipeline for the
Planck mission \citep{rein06}, that has been made available for the
Planck community by the Planck LevelS team. The \texttt{simmission}
code \citep{floor02,challinor02} has been used to simulate the
satellite dynamics during a mission flight of 14 months. The
satellite will observe the microwave sky from the Lagrangian point
L2, where it will be spin-stabilised by rotating at 1 \emph{r.p.m.}
The rotation axis will progress as the Earth moves around the Sun,
therefore Planck will cover the full sky in half a year. The angle
between its optical and rotation axes will be slightly less than
$90^{\circ}$, thus Planck will scan the sky along circles. In order
to cover the full sky, its rotation axis will make slow excursions
from the Ecliptic plane with an amplitude of several degrees.

For this simulation we have assumed a cycloidal scanning strategy
with a $7^{\circ}$ slow variation in the Ecliptic collatitude of the
z-axis. This scanning strategy implies that the sky will be covered
inhomogeneously. As mentioned before, the sky is covered twice in a
year. However, we have simulated a flight duration of 14 months (the
nominal mission duration). In the additional 2 months a region that
covers approximately $1/3$ of the total area of the sky is observed
one more time. Figure 1 shows the observation pattern, in Ecliptic
coordinates, for the 70 GHz channel. The observation pattern is
obtained from the satellite dynamics taking into account the
position of the different detectors in the focal plane, so the
observation pattern for 100 GHz is slightly different from the one
in Figure 1. In both cases we can see a broad ``ring'' where the
number of observations is higher than in the rest of the sky. This
``ring'' covers $\approx 1/3$ of the sky. Since the final noise
level is inversely proportional to the square root of the number of
observations, the regions inside that ring have lower noise levels
than the rest of the sky. In order to generate anisotropic noise
patches for our work we have taken two representative zones of the
sky, one inside the ``low noise'' ring and other outside it. The
noise rms per pixel is obtained from the number of observations
inside the pixel and the detector noise equivalent temperature, that
is specified in the Planck detector database. The final average
noise rms levels per pixel (pixel size $=3^{\prime}$) in
thermodynamic $\Delta T/T$ units are $3.51\times 10^{-5}$
($1.35\times 10^{-5}$) at 70 (100) GHz for the high noise zone and
$2.19\times 10^{-5}$ ($8.68\times 10^{-6}$) at 70 (100) GHz for the
low noise zone (inside the ring).  To preserve the proportion of
high to low noise regions in the sky we have simulated two high
noise patches for every low noise patch.

 We consider 1000 simulations of CMB plus anisotropic noise at 70 and 100 GHz,
these simulations are Gaussian and will serve as comparison with the
non-Gaussian ones, one third of the simulations correspond to the
low noise zone and two thirds to the high noise zone. Then, we
generate 1000 simulations including CMB, anisotropic noise, Galactic
foregrounds and point sources for each of the following flux limits:
1Jy, 0.9 Jy, 0.8 Jy,......., 0.1 Jy, i.e. we remove all the sources
above the given flux limit (the third order and fourth order
cumulants of the different simulated components can be seen in Table
2). We filter all the simulations with the MHW1, MHW2 and MHW3 at
the optimal scales (obtained by maximizing the amplification) and
calculate their skewness and kurtosis.

Then, taking into account the skewness, we perform the following
hypothesis test: we consider the null hypothesis, $H_0$, the
simulation is Gaussian, against the alternative hypothesis, $H_1$,
it is non-Gaussian. We set a significance level $ \alpha =0.05 $,
this means that we reject the null hypothesis when a simulation with
point sources has a skewness higher than a value $s_{5}$, which
corresponds to that of the $95\%$ of the Gaussian simulations. The
test is one-sided to the right, since the simulations with point
sources have positive skewness. We define the power of the test as
$1-\delta$, with $\delta$ the probability of accepting the null
hypothesis when it is false, i.e. the power is the proportion of
non-Gaussian simulations with a skewness higher than that of $95\%$
of the Gaussian ones. The higher the power, the
 more efficient the method in detecting non-Gaussianity.

We perform the same test with the kurtosis, setting a kurtosis limit
$k_{5}$; it is also one-sided to the right, since the kurtosis of
point sources is higher than 3.

The power of these tests for the different flux limits are shown in
Table 1, they are also plotted in Figure 2. At 70 GHz the power of
the skewness test with a flux limit of 0.4 Jy is $8\%$ (MHW1), $9\%$
(MHW2) and $10\%$ (MHW3). The power is higher for the kurtosis test
: $11\%$ (MHW1), $13\%$ (MHW2) and $11\%$ (MHW3) at 0.4 Jy.
Therefore, for the MHW2 the probability that a patch with this flux
limit has a kurtosis higher than that of $95\%$ of the Gaussian
simulations is 0.13. In practice, we could consider 126 patches
(half sky) with the size of the simulated ones, if they are Gaussian
the probability of finding more than 10 with a kurtosis higher than
$k_{5}$ is 0.05, if we had point sources up to 0.4 Jy, this
probability would be 0.95, (this can be calculated by using the
binomial distribution with probabilities 0.05 and 0.13 in the
Gaussian and non-Gaussian case, respectively). Therefore, the
detection of non-Gaussianity would be possible for this flux limit.

For the 100 GHz channel, the results of the skewness test can also
be seen in Table 1, for instance the power at 0.3 Jy is $8\%$
(MHW1), $13\%$ (MHW2) and $12\%$ (MHW3). The power is higher for the
kurtosis test: $9\%$ (MHW1), $18\%$ (MHW2), $22\%$ (MHW3) at 0.3 Jy,
making detection possible down to this flux. The kurtosis test
allows the detection of non-Gaussianity at 100 GHz for lower fluxes
than at 70 GHz.

We would like to comment that point sources could be detected in
these channels, accepting a $5\%$ of spurious detections, down to
$0.47$ Jy (70 GHz) and $0.25$ Jy (100 GHz)\citep{lop06}. Therefore,
even if we were able to perform a perfect subtraction (or masking)
of the point sources above these detection fluxes, we would still
clearly detect a non-Gaussian signal due to unsubtracted point
sources in the 70 GHz channel. For 100 GHz, at the flux cut of 0.2
Jy  (i.e below the theoretical detection limit), the highest found
power is $11\%$ (corresponding to the kurtosis of the MHW3). Thus,
it is likely that a non-Gaussian signal due to residual point
sources is still present when analysing a presumably clean CMB map.
These results show the importance of studying and characterising the
non-Gaussianity due to residual point sources (or other foregrounds)
in order to avoid its misidentification as intrinsic
non-Gaussianity.

In general the MHW2 and MHW3 perform better than the MHW1, producing
higher powers; the kurtosis test is more powerful than the skewness
one and the tests are more powerful at 100 GHz than at 70 GHz, (see
Table 1 and Figure 2).

In order to assess the influence of the Galactic foregrounds, we
have also carried out 1000 simulations without point sources but
including the Galactic foregrounds. The power in this case is always
below $6\%$ (see the no sources case in Table 1), confirming that
these foregrounds contribute very little to the non-Gaussianity,
when we filter with these wavelets at the optimal scales.

We have also performed similar tests by filtering the maps with the
Daubechies 4 wavelet, this wavelet has been already used in the
study of the non-Gaussianity of CMB maps \citep{bar01}. The power of
the skewness and kurtosis tests with this wavelet is never higher
than $50\%$, even for fluxes as high as 1 Jy and is always much
lower than the power obtained with the MHWF, as can be seen in
Figure 2. We have considered skewness and kurtosis of different
detail coefficients (horizontal, vertical, diagonal) at different
resolution levels and plotted in Figure 2 the power obtained for the
detail coefficients which produce the highest power. It is clear
that, whereas the members of the MHWF are adapted to the detection
of point sources and the non-Gaussian signal they produce, this is
not the case for the Daubechies family.

\section{Characterising point sources}

\subsection{Point source cumulants}

In the previous section, we have performed Gaussianity tests in
order to detect the non-Gaussian signal generated by point sources
in CMB maps. The estimation of the level of this signal is also of
great interest. Therefore, it is worthwhile calculating the third
order cumulant, $K_3$, and the fourth order cumulant, $K_4$,
produced by point sources. The problem is that we could only know
directly the values of $K_{3}$ and $K_{4}$ of CMB maps including
CMB, noise, point sources and other foregrounds, but not of the
point source maps. A possible solution would be to filter the maps
with suitable wavelets so that the main contribution to the
cumulants would be that of point sources. As we have seen before,
the MHWF at the optimal scales could perform well in this respect.
Therefore, our idea is to filter CMB maps with the MHWF and to try
to extract the parameters $K_3$ and $K_4$. However, we should first
relate the cumulants of a wavelet filtered map to the cumulants of
the original map, this can be done by using eqs. (14) and (15).
These equations for the MHW1, MHW2 and MHW3 give us the relation
between the original cumulants due to point sources and the wavelet
filtered cumulants.

Our goal is to estimate $K_3$ and $K_4$ from simulations including
CMB, noise, point sources and other foregrounds at 70 and 100 GHz.
We filter these simulations with the MHW1, MHW2 and MHW3 at their
optimal scales. We include two additional scales for each wavelet to
improve the fit, one arcminute less than the optimal scale and one
arcminute more than the optimal scale. So for each simulated map we
calculate nine quantities $c^i$, the third order cumulants for the
three scales and the three wavelets involved, where we have
subtracted the average contribution of the anisotropic noise. We
perform a chi-squared fit to the third order cumulant of the wavelet
filtered maps $g^{i}K_3$, with $g^i$ the coefficients, see eqs. (14)
and (15), which relate the wavelet filtered third order cumulant
with the original one $K_3$. The chi-squared function can be written

\begin{equation}
\chi^2=\sum_{i,j} C_{ij}^{-1} (c^i-g^iK_3)(c^j-g^jK_3)
\end{equation}

\noindent with $C_{ij} $ the covariance matrix of the wavelet
filtered third order cumulants obtained from 1000 simulations
including CMB plus noise. The simulations used are flat patches of
$12.8\times12.8$ deg$^2$ similar to the ones explained in section 3.
Minimizing the chi-squared expression with respect to $K_3$, we
estimate the third order cumulant due to point sources of the
original map. We carry out 126 simulations (half sky) including CMB,
anisotropic noise, point sources up to a certain flux limit (from
0.1 Jy to 1 Jy) and other foregrounds. The third order and fourth
cumulants of the different components of these simulations can be
seen in Table 2. Note that in this Table, we write the cumulants
after filtering the maps with the Gaussian instrumental beam, except
for the noise.

 We obtain the point source cumulant $K_3$ from each map through the
 chi-squared
minimization. Finally, we calculate an average $K_3$ representative
of half sky. We perform the same process for the fourth order
cumulant $K_4$. The chi-squared function is in this case

\begin{equation}
\chi^2=\sum_{i,j} D_{ij}^{-1} (d^i-h^iK_4)(d^j-h^jK_4)
\end{equation}

\noindent Now $D_{ij}$ is the covariance matrix of the fourth order
cumulants, $d^i$ the fourth order cumulants of the wavelet filtered
maps (subtracting the average anisotropic noise contribution) and
$h^i$ the coefficients which relate the wavelet filtered fourth
order cumulants and the original ones. Minimizing with respect to
$K_4$, we estimate its value for each map. As before, we calculate
an average value from 126 simulations (half sky). Finally, we
calculate the average value of $K_3$ and $K_4$ from ten half skies
and their corresponding rms deviations. We have also calculated
these cumulants directly from simulations (unfiltered) of the same
size, which only include point sources up to a certain flux limit to
compare our previous estimations with the real values. The estimated
values with the MHWF and the values obtained from the point source
only simulations, $K_3^{ps}$ and $K_4^{ps}$, are presented in Table
3 for different flux limits from 0.1 Jy to 1 Jy. We have also
calculated the relative error of $K_3$ and $K_4$ with respect to
$K_3^{ps}$ and $K_4^{ps}$ for each half sky. We write in Table 3 the
average and the dispersion of this error obtained from ten half
skies. We can see in this table that the average relative error of
$K_3$ and $K_4$ with respect to $K_3^{ps}$ and $K_4^{ps}$ is below
$12\%$ down to $0.6$ Jy (70 GHz) and $0.4$ Jy (100 GHz), so that
values close to the real ones can be obtained down to low fluxes.

Finally, in Table 4, we write $K_3^{ps}$ and $K_4^{ps}$ calculated
by carrying out point source simulations in which we use the
Toffolatti source counts model \citep{tof98} as input for our source
counts. It is clear that our estimations, $K_3$ and $K_4$, (see
Tables 3 and 4) would allow us to distinguish between the De Zotti
and the Toffolatti model.

\subsection{Fit of the counts to a power law}

Differential number counts can be modeled as a function of the flux
$S$ by a power law

\begin{equation}
\label{cc} \displaystyle\frac{dn}{dS}=A S^{-\alpha}
\end{equation}

Though the fit is not very good for all the fluxes \citep{dez05}, it
can be suitable for a certain range of fluxes, for instance for
fluxes from 0.1 Jy to 1 Jy. Being able to obtain the values of the
parameters $A$, the amplitude, and $\alpha$, the slope, can be very
useful to distinguish between different source counts models and to
predict the number counts in the corresponding flux ranges.

The third order and fourth order cumulants of the source temperature
distribution in a pixelized map with pixel area $A_p$ can be easily
calculated from the differential number counts assuming that sources
are Poissonian distributed. This is a good approximation for the
frequencies (70 and 100 GHz) we are considering \citep{gon05}. In
this case, the third order cumulant can be written

\begin{equation}
K_3=A_{p}^{-2}g^{3}(x)\int_{0}^{S_{lim}} dS \,S^{3}\,\frac{dn}{dS}
\end{equation}
\noindent where $S_{lim}$ is the flux limit and $g$ is the
conversion factor from fluxes to temperatures.
\begin{equation}
 g(x)\equiv 2\frac{(hc)^{2}}{(k_BT)^{3}}\frac{(sinh ({x}/{2}))^{2}}{x^{4}}
\end{equation}
\noindent and $x\equiv h\nu/k_BT$.

The formula for the fourth order cumulant is similar

\begin{equation}
K_4=A_{p}^{-3}g^{4}(x)\int_{0}^{S_{lim}} dS \,S^{4}\,\frac{dn}{dS}.
\end{equation}

 If we assume a power law for the differential number counts, we
could obtain $K_3$ and $K_4$ just by introducing the expression
$AS^{-\alpha}$ in the previous formulas

\begin{equation}
K_{3}=\frac{Ag^3 S_{lim}^{4-\alpha}}{(4-\alpha)A_{p}^{2}}
\end{equation}

\begin{equation}
K_{4}=\frac{Ag^4 S_{lim}^{5-\alpha}}{(5-\alpha)A_{p}^{3}}
\end{equation}

\noindent Then, A and $\alpha$ could be obtained from these
equations, assuming that we know $K_3$, $K_4$ and $S_{lim}$.

\begin{equation}
\alpha=\frac{4-5c}{1-c}
\end{equation}

\begin{equation}
A=\frac{(4-\alpha)K_{3} A_{p}^{2}}{g^{3}S_{lim}^{(4-\alpha)}}
\end{equation}

\noindent with

\begin{equation}
c=\frac{A_{p} K_{4}}{g K_{3} S_{lim}}
\end{equation}

After estimating $K_3$ and $K_4$ for the 70 and 100 GHz channels by
using the fit with the MHWF (see the previous subsection), we can
obtain the estimated values of the amplitude, $A$, and the slope,
$\alpha$ for each half sky by applying eqs. (32), (33) and (34). If
we consider ten half skies, we can give an average value and the rms
deviation.

The results can be seen in Table 5, where they are compared with
those obtained from the point source only simulations; in this last
case we have used the values $K_{3}^{ps}$ and $K_{4}^{ps}$,
calculated previously from the point source maps. The results at 70
GHz for a 1 Jy flux limit are: $\alpha=2.19\pm 0.43$, $A=24.3\pm
4.3$, whereas we obtain from the point source simulations:
$\alpha^{ps}=2.32\pm 0.06$, $A^{ps}=22.1\pm 1.5$. For this flux
limit, the average relative error of $\alpha$ ($A$) with respect to
$\alpha^{ps}$ ($A^{ps}$) is below $10\%$ ($15\%$). The estimation is
not bad, but the rms deviation is high. We can obtain reasonable
estimations down to 0.7 Jy. It is clear that small errors in the
determination of the cumulants produce high errors in the amplitude
and slope.

The values obtained from the fit for a flux limit of 1 Jy at 100 GHz
are: $\alpha=2.26\pm 0.19$, $A=21.0\pm 4.2$, whereas from the point
source simulations we estimate: $\alpha^{ps}=2.35\pm 0.16$,
$A^{ps}=20.0\pm 3.7$. The average relative error of $\alpha$
($A$)with respect to $\alpha^{ps}$ ($A^{ps}$) is below $5\%$
($6\%$). We obtain reasonable values down to 0.7 Jy, with less
dispersion than for the 70 GHz channel.

In Figure 3, the fit for both channels at 1 Jy is compared to the
differential number counts of the De Zotti model, which was used for
the simulations and with the estimation from the point source
simulations. As can be seen in the plots, the fit is quite good for
fluxes between 0.1 Jy and 1 Jy.

\section{Conclusions}

We have considered a natural generalization of the MHW1 on the
plane, $R^2$, based on the iterative application of the Laplacian
operator to the MHW1. We have called this group of wavelets the
Mexican Hat Wavelet Family (MHWF) \citep{gon06,lop06}. We have
applied the MHWF to the study of the non-Gaussianity produced in CMB
maps by extragalactic point sources, a very important issue for the
separation and analysis of different non-Gaussian components in the
CMB.

We have first derived a formula, eq. (14), which relates the
cumulants of 2D point source maps filtered with the MHWF with the
cumulants of the original point source maps. From this formula, we
have obtained several equations, eqs. (18-23), which relate the
contribution of point sources to the skewness and kurtosis of
wavelet (MHWF) filtered 2D CMB maps to their contribution to the
skewness and kurtosis of beam filtered CMB maps.

 According to these formulas, the skewness of a wavelet filtered
 map is proportional to $\lambda^{3}$, with $\lambda$ the
 amplification, and the kurtosis is proportional to $\lambda^{4}$.
 This means that the higher the amplification the higher the
 non-Gaussian signal will be. Therefore, the MHW2 and MHW3 at their optimal scales
 can be very suitable filters to detect the non-Gaussian signal due
 to point sources, since they produce a higher amplification than
 the MHW1 \citep{gon06,lop06}.

 We have then considered Gaussian simulations of CMB plus
 anisotropic noise and compared them with simulations including all the
 relevant components: CMB, anisotropic noise, point sources and Galactic
  foregrounds. The simulations are 2D maps of $12.8\times12.8$ square degrees
  with the characteristics of the Planck experiment at 70 GHz and
  100 GHz. We perform a Gaussianity test based on the skewness and
  kurtosis of the maps filtered with the MHW1, MHW2 and MHW3 at the
  optimal scales (those which produce the highest amplifications),
   we compare 1000 simulations of CMB plus noise
   with 1000 simulations including point sources up to different flux limits from 0.1 Jy to
   1 Jy.

  The power of the test is higher for the MHW2 and the MHW3 than for
  the standard MHW1, the kurtosis test is more powerful than the skewness test.
  We could detect the non-Gaussianity due to point sources down to a
  flux limit of 0.4 Jy (70 GHz) and 0.3 Jy (100 GHz), the powers are higher
  for 100 GHz than for 70 GHz. The results can be seen in Figure 2 and
Table 1. We would like to point out that according to \cite{lop06}
point sources could be detected in these channels, assuming a $5\%$
of spurious detections, down to $0.47$ Jy (70 GHz)and $0.25$ Jy (100
GHz). This means that, using the MHWF, we can see a non-Gaussian
signal due to weak point sources, which are in the limit of
detection of current source extraction techniques. Therefore, it is
very important to characterise the non-Gaussianity due to
unsubtracted point sources, in order to avoid its possible
misidentification as intrinsic non-Gaussianity.

The influence of Galactic foregrounds is negligible, since when we
consider simulations without point sources, but keeping the Galactic
foregrounds, the powers go down below $6\%$. We have also studied
the performance of the Daubechies 4 wavelet, obtaining powers below
$50\%$ even for flux limits as high as 1 Jy. The MHWF performs much
better than this wavelet (see Figure 2), since it is adapted to deal
with point sources.

This method could be used for all the frequencies of the Planck
experiment and could also be extended to spherical CMB images, as
the ones provided by the Planck experiment.

We have also tried to estimate the third order cumulant $K_3$ and
the fourth order cumulant $K_4$, produced by point sources at the 70
and 100 GHz Planck channels. We filter $12.8\times 12.8$ square
degrees CMB maps with the characteristics of the Planck experiment
with the MHW1, MHW2 and MHW3 at three scales for each wavelet
(including the optimal ones). The maps include CMB, anisotropic
noise, point sources and other foregrounds. We calculate the third
and fourth order cumulants of these wavelet filtered maps and write
chi-squared functions, eqs. (24) and (25), which allow us to
estimate the cumulants of the original point source maps; we have
used eq. (14) to relate the cumulants of the wavelet filtered maps
with those of the original maps. We have compared our estimated
values with those, $K_3^{ps}$ and $K_4^{ps}$, obtained directly from
simulations of the same type but including only unfiltered point
sources without any other component. The results can be seen in
Table 3. The average error (from ten half skies) is below $12\%$
down to a flux limit of $0.6$ Jy (70 GHz) and $0.4$ Jy (100 GHz).
The determination of these cumulants would allow us to distinguish
between different source counts models (see Tables 3 and 4).

Finally, we have also studied the fit of the differential number
counts at 70 and 100 GHz to a power law $dn/dS=A S^{-\alpha}$.
 From the cumulants previously estimated, we calculate by
using eqs. (32-34) the amplitude $A$ and slope $\alpha$ of the
differential number counts. The results at 70 GHz for a 1 Jy flux
limit are: $\alpha=2.19\pm 0.43$, $A=24.3\pm 4.3$. When we calculate
these magnitudes from the cumulants $K_3^{ps}$ and $K_4^{ps}$,
computed directly from the point source only simulations, we obtain
$\alpha^{ps}=2.32\pm 0.06$, $A^{ps}=22.1\pm 1.5$. These are the
average and rms deviation from ten half skies. For this flux limit,
the average relative error of $\alpha$ ($A$) with respect to
$\alpha^{ps}$ ($A^{ps}$) is below $10\%$ ($15\%$). The dispersion in
the values obtained with the MHWF fit is high, though the average is
consistent with the value obtained directly from point sources down
to a flux limit of 0.7 Jy, as can be seen in Table 5.

When we consider the 100 GHz channel, the values obtained from the
MHWF fit for a flux limit of 1 Jy are: $\alpha=2.26\pm 0.19$,
$A=21.0\pm 4.2$, the results from the point source simulations are:
$\alpha^{ps}=2.35\pm 0.16$, $A^{ps}=20.0\pm 3.7$. The fit is better
for this channel than for the 70 GHz one, having a lower dispersion.
For this flux limit, the average relative error of $\alpha$ ($A$)
with respect to $\alpha^{ps}$ ($A^{ps}$) is below $5\%$ ($6\%$). The
results are consistent down to a flux limit of 0.7 Jy. The detailed
values of the parameters $\alpha$ and $A$ can be seen in Table 5.

In Figure 3 we plot the differential source counts of the De Zotti
model, the fit to a power law by using the MHWF and the fit obtained
from the point source only simulations, in both cases with a 1 Jy
flux limit. As we can see in this figure, both fits are consistent
with the input counts in the range between 0.1 Jy and 1 Jy.

\section*{Acknowledgments}

We acknowledge partial financial support from the Spanish Ministry
of Education (MEC) under project ESP2004--07067--C03--01. We thank
G. De Zotti for having kindly provided us with the source number
counts foreseen by the De Zotti et al. (2005) cosmological evolution
model at the LFI frequencies. We acknowledge the use of the Planck
Reference Sky, prepared by the members of the Planck Working Group
2. We also thank L. Toffolatti for useful discussions. Finally, we
thank the referee, Mark Ashdown, for his useful comments and
suggestions.

\clearpage

\begin{table*}
  \centering
  \begin{tabular}{ccccccc}
  \textbf{70 GHz}& \textbf{s(MHW1)}& \textbf{s(MHW2)} & \textbf{s(MHW3)}& \textbf{k(MHW1)}& \textbf{k(MHW2)}& \textbf{k(MHW3)}\\
  \hline\hline
  \textbf{no sources} &4 & 4 & 4 & 5 & 4 & 4 \\
    \textbf{0.1 Jy} & 5  & 5  & 5  & 5  & 5  & 5  \\
    \textbf{0.2 Jy} &5  & 6  & 6  & 6  & 6  & 6  \\
    \textbf{0.3 Jy} &7 &7  &7  &7  &8  &7   \\
    \textbf{0.4 Jy} &8  &9  &10  &11  &13  &11  \\
    \textbf{0.5 Jy} &10  &11 &10  &13  &16  &14  \\
    \textbf{0.6 Jy} &14  &17  &16  &21  &26  &23  \\
    \textbf{0.7 Jy} &21  &24  &22  &35  &39  &37  \\
    \textbf{0.8 Jy} &23  &27  &26  &39  &43  &41  \\
    \textbf{0.9 Jy} &36  &39  &37  &53  &56  &53  \\
    \textbf{1 Jy} &33  &38  &36  &52  &55  &52  \\
 \hline\hline
 \textbf{100 GHz}& \textbf{s(MHW1)}& \textbf{s(MHW2)} & \textbf{s(MHW3)}& \textbf{k(MHW1)}& \textbf{k(MHW2)}& \textbf{k(MHW3)}\\
  \hline\hline
  \textbf{no sources} & 3 & 5 & 4 & 4 & 4 & 6 \\
    \textbf{0.1 Jy} & 4  & 5  & 6  & 4  & 5  & 5  \\
    \textbf{0.2 Jy} &6  & 9  & 9  & 6  & 8  & 11  \\
    \textbf{0.3 Jy} &8 &13  &12  &10  &18  &22   \\
    \textbf{0.4 Jy} &21  &29  &30  &32  &49  &54  \\
    \textbf{0.5 Jy} &25  &35 &33  &40  &58  &61  \\
    \textbf{0.6 Jy} &39  &52  &51  &60  &75  &76  \\
    \textbf{0.7 Jy} &55  &65  &64  &72  &81  &83  \\
    \textbf{0.8 Jy} &56  &67  &66  &75  &84  &85  \\
    \textbf{0.9 Jy} &68  &76  &76  &80  &87  &89  \\
    \textbf{1 Jy} &67  &75  &74  &82  &88  &89  \\
 \hline\hline
\end{tabular}

  \caption{Power (percentage) of the skewness and kurtosis tests at 70 and 100 GHz for different flux limits, for the first three members of the
  MHWF. In the first, second and third columns, we show the power of the skewness test for the MHW1, MHW2 and MHW3, respectively.
  In the fourth, fifth and sixth columns, we write the power of the kurtosis test for the MHW1, MHW2 and MHW3.
  }\label{table1}
\end{table*}

\begin{table*}
  \centering
  \begin{tabular}{ccccc}
\hline & \multicolumn{2}{c}{\textbf{$K_{3}$}}&
\multicolumn{2}{c}{\textbf{$K_{4}$}}\\
\hline\hline Freq(GHz)& 70 & 100 & 70 & 100 \\
noise(high)& (0.07$\pm$5.9)$\times 10^{-17}$ & $(-0.1\pm$2.1)$\times
10^{-18}$ & (9.6$\pm$0.1)$\times 10^{-19}$ & (6.9$\pm$0.1)$\times
10^{-21}$
\\
noise(low) & (-0.3$\pm$1.9)$\times 10^{-17}$ & (-0.1$\pm$1.1)$\times
10^{-18}$& (6.8$\pm$0.1)$\times 10^{-20}$&(1.43$\pm$0.02)$\times
10^{-21}$\\
CMB & (-0.4$\pm$3.2)$\times 10^{-16}$ &  (-0.5$\pm$3.7)$\times
10^{-16}$&(-2.5$\pm$1.8)$\times 10^{-20}$ & (-2.3$\pm$1.9)$\times
10^{-20}$\\
Gal.fore.& (-4.5$\pm$0.6)$\times 10^{-17}$& (-4.0$\pm$0.5)$\times
10^{-17}$& (1.5$\pm$0.3)$\times 10^{-21}$& (1.4$\pm$0.3)$\times
10^{-21}$\\
1 Jy & (2.4$\pm$0.1)$\times 10^{-16}$&(1.8$\pm$0.1)$\times
10^{-16}$&(1.6$\pm$0.1)$\times 10^{-20}$&(1.4$\pm$0.1)$\times
10^{-20}$\\
0.5 Jy &  (6.1$\pm$0.2)$\times 10^{-17}$&(4.9$\pm$0.2)$\times
10^{-17}$&(2.1$\pm$0.1)$\times 10^{-21}$&(2.0$\pm$0.1)$\times
10^{-21}$\\
0.1 Jy &  (2.1$\pm$0.1)$\times 10^{-18}$&(1.60$\pm$0.04)$\times
10^{-18}$&(1.41$\pm$0.05)$\times 10^{-23}$&(1.29$\pm$0.04)$\times
10^{-23}$\\
 \hline\hline
\end{tabular}
\caption{Third order cumulant $K_3$ and fourth order cumulant $K_4$
  of the different components of our simulations: noise (high noise
  zone), noise (low noise zone), CMB, Galactic foregrounds and point
  sources, up to flux limits of 1 Jy, 0.5 Jy and 0.1 Jy. The cumulants are
  obtained from the simulations filtered with the Gaussian beam,
  except for the noise.
  We show the average and rms deviation
   from ten half skies.
  }\label{table2}
\end{table*}

\clearpage

\begin{table*}
  \centering
  \begin{tabular}{ccccccc}

  \textbf{70 GHz}& \textbf{$K_{3}$}& \textbf{$K_3^{ps}$} &\textbf{$\Delta K_{3}$} & \textbf{$K_{4}$}& \textbf{$K_{4}^{ps}$} &\textbf{$\Delta K_{4}$}\\
  \hline\hline
      \textbf{1 Jy} & (4.60$\pm$ 0.59)$\times 10^{-13}$  & (4.45$\pm$0.20)$\times 10^{-13}$& 9.5$\pm$4.9  & (1.02$\pm$0.12)$\times 10^{-15}$  & (0.97$\pm$0.05)$\times 10^{-15}$& 7.4$\pm$6.9  \\
      \textbf{0.9 Jy} & (4.29$\pm$0.38)$\times 10^{-13}$  & (4.16$\pm$0.26)$\times 10^{-13}$ & 5.4$\pm$4.4 & (9.02$\pm$1.07)$\times 10^{-16}$  & (8.86$\pm$0.70)$\times 10^{-16}$& 6.7$\pm$3.4  \\
 \textbf{0.8 Jy} & (3.50$\pm$0.27)$\times 10^{-13}$  & (3.45$\pm$0.22)$\times 10^{-13}$ & 6.1$\pm$5.5 & (6.76$\pm$0.72)$\times 10^{-16}$  & (6.63$\pm$0.54)$\times 10^{-16}$& 6.2$\pm$4.7   \\
 \textbf{0.7 Jy} & (2.68$\pm$0.31)$\times 10^{-13}$  & (2.67$\pm$0.16)$\times 10^{-13}$ & 9.8$\pm$4.9 & (4.22$\pm$0.35)$\times 10^{-16}$  & (4.44$\pm$0.32)$\times 10^{-16}$& 5.8$\pm$4.9  \\
\textbf{0.6 Jy} & (1.73$\pm$0.26)$\times 10^{-13}$  & (1.73$\pm$0.05)$\times 10^{-13}$ & 11.5$\pm$10.4 & (2.38$\pm$0.26)$\times 10^{-16}$  & (2.37$\pm$0.07)$\times 10^{-16}$& 7.8$\pm$5.8  \\
\textbf{0.5 Jy} & (1.13$\pm$0.24)$\times 10^{-13}$  & (1.12$\pm$0.04)$\times 10^{-13}$ & 18.0$\pm$10.0 & (1.14$\pm$0.42)$\times 10^{-16}$  & (1.24$\pm$0.06)$\times 10^{-16}$ &26.0$\pm$23.1  \\
\textbf{0.4 Jy} & (8.37$\pm$2.85)$\times 10^{-14}$  & (8.80$\pm$0.42)$\times 10^{-14}$ & 25.2$\pm$18.1 & (9.54$\pm$3.03)$\times 10^{-17}$  & (8.56$\pm$0.56)$\times 10^{-17}$& 26.7$\pm$23.2  \\
\textbf{0.3 Jy} & (6.44$\pm$2.24)$\times 10^{-14}$  & (4.33$\pm$0.19)$\times 10^{-14}$ & 58.4$\pm$46.7  & (4.26$\pm$2.58)$\times 10^{-17}$  & (3.03$\pm$0.16)$\times 10^{-17}$& 77.6$\pm$61.7  \\
\textbf{0.2 Jy} & (1.74$\pm$3.15)$\times 10^{-14}$  & (2.08$\pm$0.08)$\times 10^{-14}$ & 120$\pm$86 & (9.89$\pm$17.71)$\times 10^{-18}$  & (1.02$\pm$0.05)$\times 10^{-17}$ & 134$\pm$110  \\
\textbf{0.1 Jy} & (1.09$\pm$3.11)$\times 10^{-14}$  & (3.90$\pm$0.10)$\times 10^{-15}$ & 586$\pm$538 & (7.90$\pm$34.6)$\times 10^{-18}$  & (8.56$\pm$0.29)$\times 10^{-19}$& 3350$\pm$2037  \\
       \hline\hline
 \textbf{100 GHz}& \textbf{$K_{3}$}& \textbf{$K_3^{ps}$}  &\textbf{$\Delta K_{3}$} & \textbf{$K_{4}$}& \textbf{$K_4^{ps}$} &\textbf{$\Delta K_{4}$}\\
  \hline\hline
      \textbf{1 Jy} & (6.90$\pm$0.54)$\times 10^{-14}$  & (6.98$\pm$0.40)$\times 10^{-14}$ & 2.3$\pm$1.7  & (8.45$\pm$0.76)$\times 10^{-17}$  & (8.38$\pm$0.60)$\times 10^{-17}$ & 2.5$\pm$2.3  \\
      \textbf{0.9 Jy} & (6.39$\pm$0.50)$\times 10^{-14}$  & (6.54$\pm$0.41)$\times 10^{-14}$& 2.8$\pm$2.4   & (7.42$\pm$0.69)$\times 10^{-17}$  & (7.50$\pm$0.64)$\times 10^{-17}$ & 2.5$\pm$1.6  \\
 \textbf{0.8 Jy} & (5.38$\pm$0.59)$\times 10^{-14}$  & (5.50$\pm$0.42)$\times 10^{-14}$& 4.6$\pm$3.2   & (5.90$\pm$0.65)$\times 10^{-17}$  & (5.83$\pm$0.58)$\times 10^{-17}$ & 3.6$\pm$2.3  \\
 \textbf{0.7 Jy} & (4.29$\pm$0.32)$\times 10^{-14}$  & (4.25$\pm$0.24)$\times 10^{-14}$ & 4.3$\pm$2.7   & (3.86$\pm$0.29)$\times 10^{-17}$  & (3.88$\pm$0.29)$\times 10^{-17}$ & 1.9$\pm$1.5  \\
\textbf{0.6 Jy} & (2.80$\pm$0.14)$\times 10^{-14}$  & (2.99$\pm$0.17)$\times 10^{-14}$ & 6.0$\pm$5.3  & (2.36$\pm$0.16)$\times 10^{-17}$  & (2.31$\pm$0.17)$\times 10^{-17}$ & 4.6$\pm$3.4  \\
\textbf{0.5 Jy} & (1.75$\pm$0.13)$\times 10^{-14}$  & (1.88$\pm$0.09)$\times 10^{-14}$ & 7.2$\pm$7.8  & (1.23$\pm$0.07)$\times 10^{-17}$  & (1.17$\pm$0.07)$\times 10^{-17}$ & 6.1$\pm$3.8  \\
\textbf{0.4 Jy} & (1.31$\pm$0.20)$\times 10^{-14}$  & (1.43$\pm$0.10)$\times 10^{-14}$ & 11.3$\pm$7.4  & (7.97$\pm$0.41)$\times 10^{-18}$  & (7.79$\pm$0.59)$\times 10^{-18}$ & 4.0$\pm$2.7  \\
\textbf{0.3 Jy} & (6.17$\pm$1.17)$\times 10^{-15}$  & (7.19$\pm$0.27)$\times 10^{-15}$ & 16.6$\pm$14.9  & (3.35$\pm$0.39)$\times 10^{-18}$  & (2.81$\pm$0.13)$\times 10^{-18}$ & 22.8$\pm$11.1  \\
\textbf{0.2 Jy} & (1.97$\pm$1.96)$\times 10^{-15}$  & (3.39$\pm$0.09)$\times 10^{-15}$ & 57.2$\pm$40.4  & (1.52$\pm$0.25)$\times 10^{-18}$  & (0.93$\pm$0.03)$\times 10^{-18}$ & 63.4$\pm$27.2  \\
\textbf{0.1 Jy} & (-8.30$\pm$15.4)$\times 10^{-16}$  & (6.10$\pm$0.16)$\times 10^{-16}$ & 282$\pm$187  & (5.43$\pm$3.85)$\times 10^{-19}$  & (7.57$\pm$0.26)$\times 10^{-20}$ & 622$\pm$476  \\
   \hline\hline
\end{tabular}

  \caption{ First and fourth columns: third order cumulant $K_3$ and fourth order cumulant $K_4$ estimated from the
   CMB maps by carrying out a fit with the MHWF, the average from ten half skies and the rms deviation are written. Second and fifth
   columns: third order cumulant $K_3^{ps}$ and fourth order cumulant $K_4^{ps}$ estimated from the
   the point source simulations (average and rms deviation from ten half
   skies). Third and sixth columns: relative error (percentage) of $K_3$ and
   $K_4$ with respect to $K_3^{ps}$ and $K_4^{ps}$ (average and rms deviation
   from ten half skies).
  }\label{table3}
\end{table*}

\begin{table*}
  \centering
  \begin{tabular}{ccc}

  \textbf{70 GHz}& \textbf{$K_3^{ps}$}& \textbf{$K_4^{ps}$}\\
  \hline\hline
      \textbf{1 Jy} & (8.6$\pm$0.4)$\times 10^{-13}$  &(1.8$\pm$0.1)$\times10^{-15}$ \\
        \textbf{0.9 Jy} & (8.1$\pm$0.3)$\times 10^{-13}$  & (1.6$\pm$0.1)$\times10^{-15}$ \\
\textbf{0.8 Jy} & (7.1$\pm$0.2)$\times 10^{-13}$  &(1.32$\pm$0.06)$\times10^{-15}$ \\
 \textbf{0.7 Jy}& (5.4$\pm$0.2)$\times 10^{-13}$  &(8.7$\pm$0.5)$\times10^{-16}$ \\
        \textbf{0.6 Jy}& (3.9$\pm$0.1)$\times 10^{-13}$  &(5.2$\pm$0.2)$\times10^{-16}$ \\
 \textbf{0.5 Jy}& (2.60$\pm$0.06)$\times 10^{-13}$  &(2.8$\pm$0.1)$\times10^{-16}$ \\
\textbf{0.4 Jy}& (2.02$\pm$0.05)$\times 10^{-13}$  &(1.90$\pm$0.07)$\times10^{-16}$ \\
        \textbf{0.3 Jy}& (1.07$\pm$0.02)$\times 10^{-13}$  &(7.3$\pm$0.2)$\times10^{-17}$ \\
 \textbf{0.2 Jy}& (5.4$\pm$0.1)$\times 10^{-14}$  &(2.59$\pm$0.06)$\times10^{-17}$ \\
\textbf{0.1 Jy}& (1.17$\pm$0.01)$\times 10^{-14}$  &(2.55$\pm$0.03)$\times10^{-18}$ \\
 \hline\hline
 \textbf{100 GHz}& \textbf{$K_{3}^{ps}$}& \textbf{$K_4^{ps}$}\\
  \hline\hline
      \textbf{1 Jy} & (1.49$\pm$0.07)$\times 10^{-13}$  &(1.7$\pm$0.1)$\times10^{-16}$ \\
        \textbf{0.9 Jy} & (1.4$\pm$0.1)$\times 10^{-13}$  & (1.6$\pm$0.1)$\times10^{-16}$ \\
\textbf{0.8 Jy} & (1.20$\pm$0.04)$\times 10^{-13}$  &(1.24$\pm$0.06)$\times10^{-16}$ \\
 \textbf{0.7 Jy}& (9.1$\pm$0.3)$\times 10^{-14}$  &(8.1$\pm$0.4)$\times10^{-17}$ \\
        \textbf{0.6 Jy}& (6.6$\pm$0.2)$\times 10^{-14}$  &(5.0$\pm$0.2)$\times10^{-17}$ \\
 \textbf{0.5 Jy}& (4.3$\pm$0.1)$\times 10^{-14}$  &(2.6$\pm$0.1)$\times10^{-17}$ \\
\textbf{0.4 Jy}& (3.4$\pm$0.1)$\times 10^{-14}$  &(1.81$\pm$0.05)$\times10^{-17}$ \\
        \textbf{0.3 Jy}& (1.81$\pm$0.03)$\times 10^{-14}$  &(6.8$\pm$0.1)$\times10^{-18}$ \\
 \textbf{0.2 Jy}& (9.3$\pm$0.1)$\times 10^{-15}$  &(2.46$\pm$0.04)$\times10^{-18}$ \\
\textbf{0.1 Jy}& (1.98$\pm$0.02)$\times 10^{-15}$  &(2.41$\pm$0.03)$\times10^{-19}$ \\
 \hline\hline

  \hline\hline
\end{tabular}

  \caption{ Third order cumulant $K_3^{ps}$ and fourth order cumulant $K_4^{ps}$ calculated from the
   the point source simulations carried out with the source number counts of the Toffolatti et al. model (average and rms deviation from ten half
   skies).
  }\label{table4}
\end{table*}

\begin{table*}
  \centering
  \begin{tabular}{ccccccc}

  \textbf{70 GHz}& \textbf{$\alpha$}& \textbf{$\alpha$$^{ps}$} &  \textbf{$\Delta \alpha$} &\textbf{A}& \textbf{A$^{ps}$}& \textbf{$\Delta$A}\\
  \hline\hline
      \textbf{1 Jy} & 2.19$\pm$0.43  & 2.32$\pm$0.06 & 9.9$\pm$16.1 & 24.3$\pm$4.3  & 22.1$\pm$1.5 & 14.9$\pm$14.9 \\
    \textbf{0.9 Jy} &2.08$\pm$0.50  & 2.10$\pm$0.12 & $15.5\pm$13.4 & 28.3$\pm$9.4  & 27.0$\pm$3.6 & 18.9$\pm$17.2 \\
    \textbf{0.8 Jy} &1.81$\pm$0.72 &2.03$\pm$0.12 & 27.2$\pm$21.1 &28.3$\pm$9.4  &27.0$\pm$3.6 & 44.6$\pm$43.3  \\
    \textbf{0.7 Jy} &2.24$\pm$0.46  &2.06$\pm$0.05 & 18.6$\pm$12.3 &25.2$\pm$9.4  &28.8$\pm$2.6 & 26.5$\pm$15 \\
     \hline\hline
 \textbf{100 GHz}& \textbf{$\alpha$}& \textbf{$\alpha$$^{ps}$} &  \textbf{$\Delta \alpha$} &\textbf{A}& \textbf{A$^{ps}$}& \textbf{$\Delta$A}\\
  \hline\hline
    \textbf{1 Jy} & 2.26$\pm$0.19  & 2.35$\pm$0.16 & 4.2$\pm$3.5 & 21.0$\pm$4.2  & 20.0$\pm$3.7 & 5.6$\pm$5.9\\
    \textbf{0.9 Jy} &2.16$\pm$0.14  & 2.23$\pm$0.13 & 4.3$\pm$3.8 & 23.4$\pm$3.4  & 22.9$\pm$3.2 & 5.7$\pm$4.9 \\
    \textbf{0.8 Jy} &1.84$\pm$0.26 &2.07$\pm$0.12 & 10.9$\pm$13.7 &30.6$\pm$5.7  &26.5$\pm$4.1 & 17.5$\pm$22.1   \\
    \textbf{0.7 Jy} &1.93$\pm$0.16  &2.14$\pm$0.11 & 10.0$\pm$6.6 &28.6$\pm$3.7  &24.8$\pm$3.8 & 16.4$\pm$12.6\\
   \hline\hline
\end{tabular}

  \caption{ First and fourth columns: slope $\alpha$ and amplitude $A$ of the differential number counts estimated from the
   CMB maps by carrying out a fit with the MHWF. Second and fifth columns: slope $\alpha^{ps}$ and amplitude $A^{ps}$ estimated from the point source simulations.
 Third and sixth columns: relative error (percentage) of $\alpha$ and
   $A$ with respect to $\alpha^{ps}$ and $A^{ps}$. In all the cases, we write the average and rms deviation
   from ten half skies }\label{table4}
\end{table*}

\clearpage
\begin{figure}
  \begin{center}
    \includegraphics[width=8.5cm]{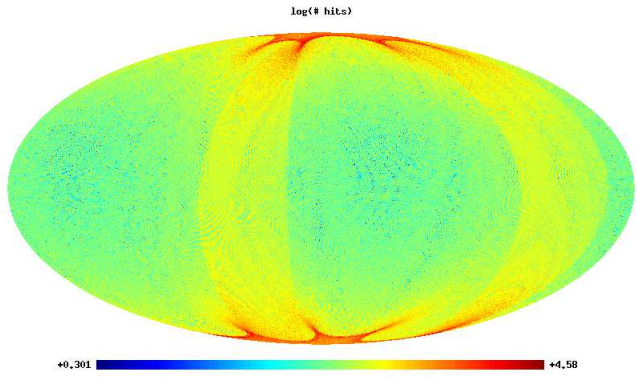}
    \caption{\label{fig1}{\small Observation pattern for the 70 GHz channel. The map shows the logarithm
of the number of hits (number of times the detector observes a given
pixel during the flight) in Ecliptic coordinates.}}

  \end{center}
\end{figure}

\clearpage
\begin{figure*}
\epsfxsize=200mm
 \epsffile{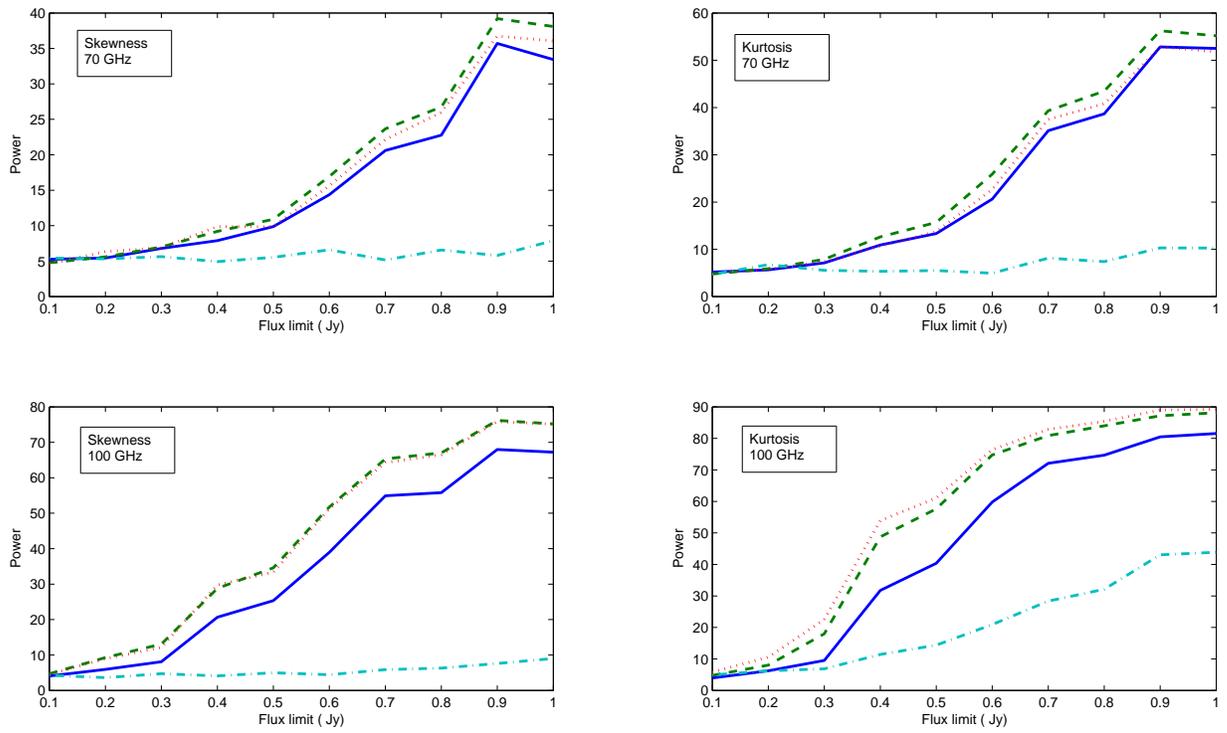} \caption{\label{fig2}{\small
 Power against the flux limit for the skewness test at 70 GHz (top
left panel), the kurtosis test at 70 GHz (top right panel), the
skewness test at 100 GHz (bottom left panel) and the kurtosis test
at 100 GHz (bottom right panel). The solid line represents the MHW1
results, the dashed line the MHW2 results, the dotted line the MHW3
results and the dash-dotted line the Daubechies4 results.}}
\end{figure*}

\clearpage

\begin{figure*}
\epsfxsize=200mm
 \epsffile{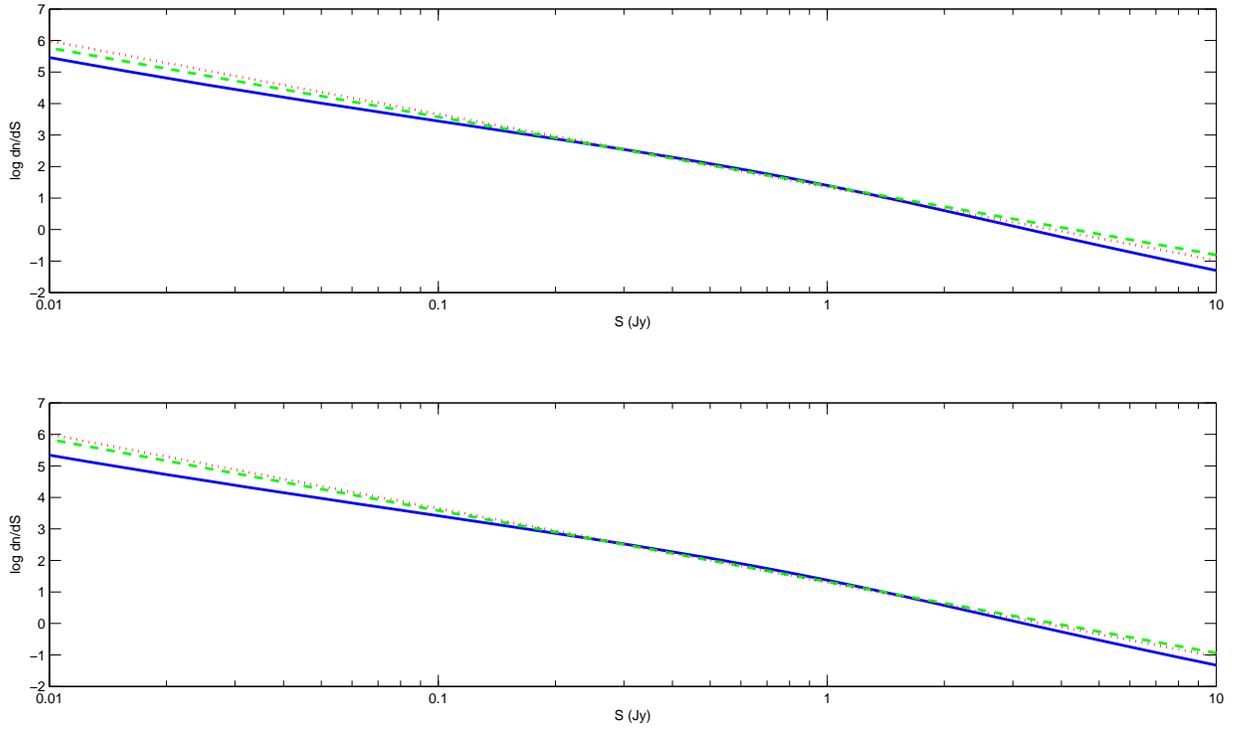} \caption{\label{fig3}{\small
 Top panel: (70 GHz channel) differential number counts against the
flux in Jy according to the De Zotti model (solid line), the fit to
a power law by using the MHWF (dashed line) with the average value
of the parameters for a flux limit of 1 Jy and the same fit by using
the original point source maps (dotted line) with the average value
of the parameters for a flux limit of 1 Jy. Bottom panel: the same
for the 100 GHz channel.}}
\end{figure*}

\end{document}